\documentstyle[11pt,iau185,twoside,epsf]{article}

\begin{document}

\title{The influence of dust properties on the mass loss in pulsating AGB stars}

\author{Anja C.\,Andersen, Susanne H\"{o}fner, Rita Loidl}

\affil{Department of Astronomy \& Space Physics, Uppsala University, Box 515, SE-751\,20 Uppsala, Sweden}

\begin{abstract} 
We are currently
studying carbon based dust types of relevance for carbon-rich AGB
stars, to obtain a better understanding of
the influence of the optical and chemical properties of the grains
on the mass loss of the star.  
An investigation of the complex interplay between hydrodynamics,
radiative transfer and chemistry has to be based on a better knowledge
of the micro-physics of the relevant dust species. 
\end{abstract}

\section{Introduction}

Asymptotic giant branch (AGB) stars show large amplitude
pulsations with periods of about 100 to 1000 days.
The pulsation creates strong shock waves in the
stellar atmos\-phere, causing a levitation of the outer layers. This cool
and relatively dense environment provides favorable conditions for the
formation of molecules and dust grains. The dust formation basically 
determines the mass loss in these stars. 
Dust grains therefore play a very important
role for the further evolution of the star (Sedlmayr this volume).

Dust formation can only take place if (1) the temperature is sufficiently low, (2) 
the abundance of the dust forming species is sufficiently large and (3) the time scale 
providing favorable conditions is sufficiently long to allow for effective
dust formation to proceed. 

Condensation and evaporation of dust in envelopes of pulsating stars
must be treated as a time-dependent process since the time scales for
condensation and evaporation are comparable to variations of the
thermodynamic conditions in the stellar atmosphere. The radiation pressure on
newly formed dust grains can enhance excisting shock waves or even create shock
waves leading to more or less pronounced discrete dust shells in the expanding
circumstellar flow (e.g.\ Fleischer et al.\ 1992; H\"ofner \& Dorfi
1997). We have calculated models of carbon-rich AGB stars with
different carbon dust properties, in order to establish the
dependence of the dynamical models on the material properties such as 
the opacity and the intrinsic density of the dust material.

\section{Carbon grains}

Amorphous carbon grains seem to be a very good candidate as the most common
type of dust particles present in circumstellar envelopes of carbon-rich AGB stars.
 
\begin{table}
\caption{List of the different dust data shown in Fig.\,1.}
\begin{center}
\begin{tabular}{|l|c|c|c|c|l|} \hline
Reference & Material & $\rho$ & Designation &  Comments \\
 & name & (g/cm$^{3}$) &  in this paper & \\ \hline
J\"{a}ger et al.\ (1998) & cel400  & 1.435 &  J\"{a}ger~400 & ``diamond-like´´ \\ 
J\"{a}ger et al.\ (1998) & cel1000 & 1.988 &  J\"{a}ger~1000 & ``graphite-like´´ \\
Maron (1990) & AC2 & 1.85 &  Maron & {\it a} \\ \hline
\end{tabular}
\end{center}
{$^a$}{\small Optical constants based on measurements by Bussoletti et al.\ (1987).} \\
\end{table}
 
There exists a wide variety of possible amorphous carbon grain types,
which fall in between the categories ``diamond-like'' and
``graphite-like'' amorphous carbon depending on the dominant type of chemical bonds. 
Different amorphous carbon dust data are listed in Table 1.  
The extinction efficiency data presented
in this paper were calculated in the Rayleigh approximation for spheres (see Andersen et al.\ 
(1999) for details). As can be seen in Fig.\,1a the difference
in optical properties of different types of amorphous carbon is substantial.

\section{Dynamical models}
 
To obtain the structure of the stellar atmos\-phere and circumstellar envelope
as a function of time
we solve the coupled system of frequency-dependent 
radiation hydrodynamics and time-dependent
dust formation (see  H\"ofner 1999 and H\"ofner et al. this volume for details). 
The dust formation is treated by the so-called moment method
(Gail \& Sedlmayr 1988; Gauger et al.\ 1990).
In the moment method dust formation is regarded as a two step process;
(1) the formation of supercritical nuclei out of the gas phase 
and (2) the time dependent growth of grains to macroscopic sizes. 
The moment method is concerned with the time evolution of an ensemble of dust
grains of various sizes and requires the nucleation rate as external input. 

\begin{figure}
 \plottwo{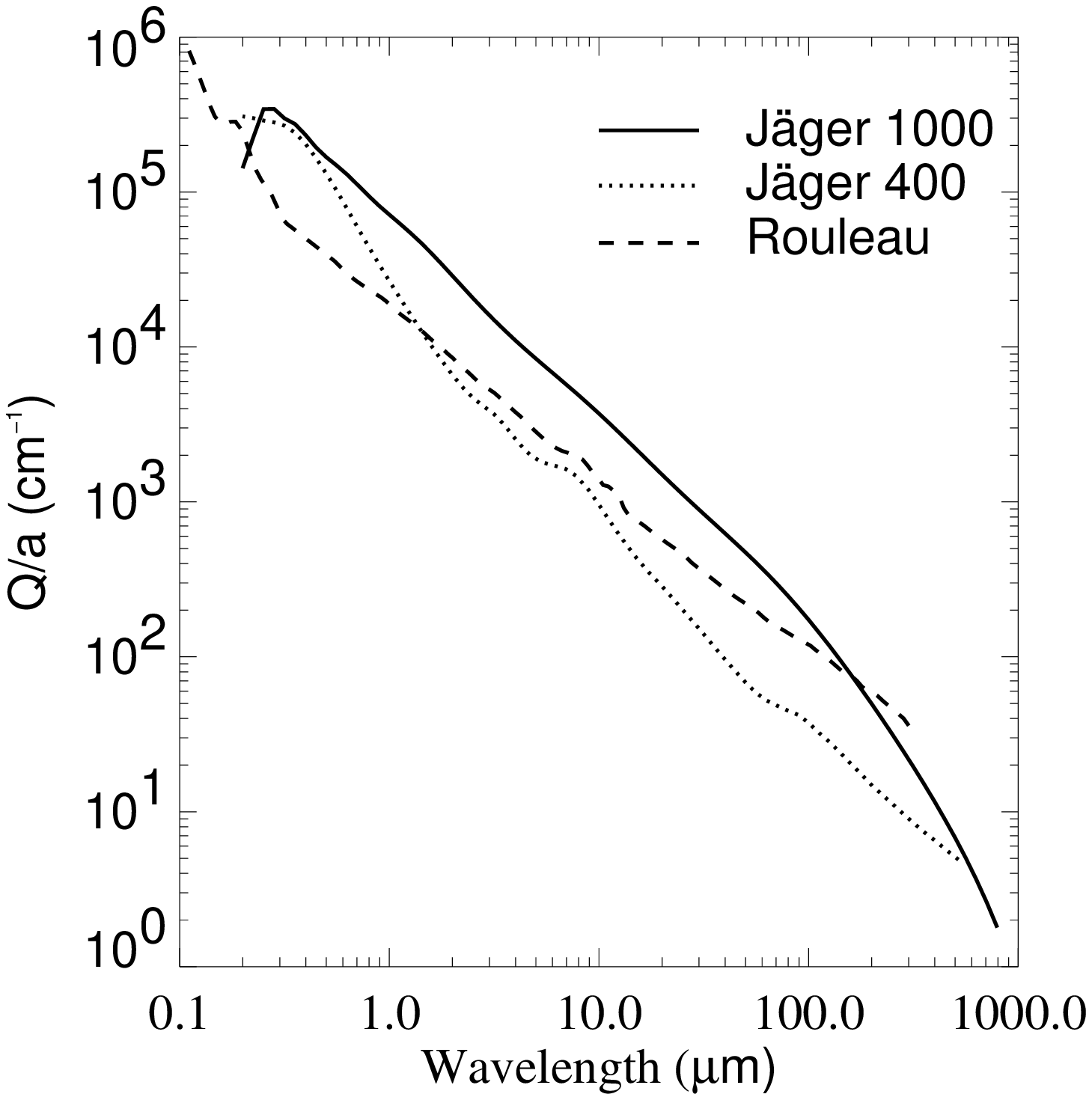}{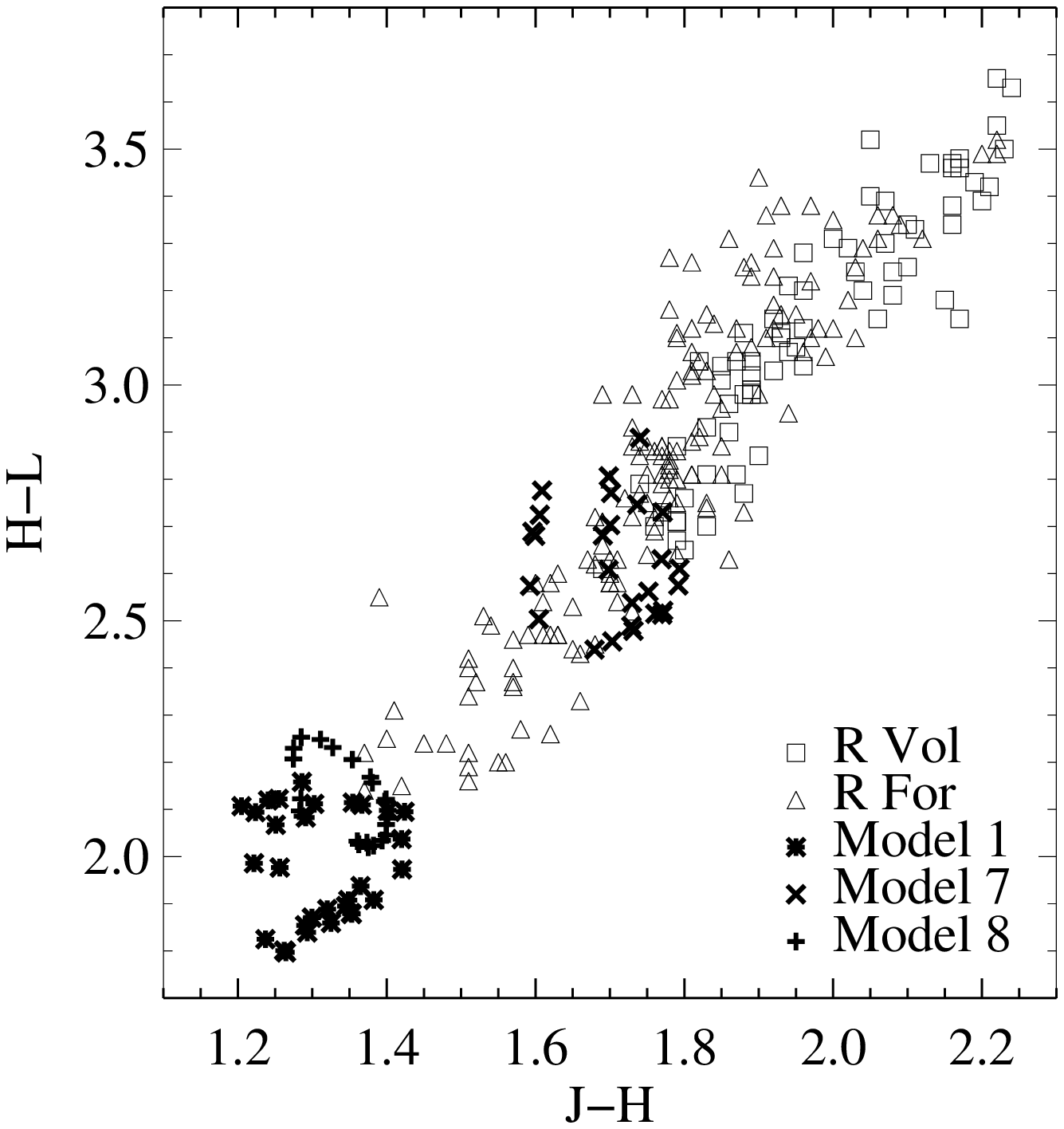}
 \caption{The figure (1a) to the left shows the extinction efficiency of amorphous carbon derived
from optical constants (see Table 1 for annotations).
The figure to the right (1b) shows the (H$-$L) vs. (J$-$H) colors for two different models 
compared to observations of the bright Mira stars \object{R\,Vol} and \object{R\,For} which have 
moderate dust shell (Whitelock et al. 1997).}
\end{figure}
 
The models require as input the extinction efficiency $Q_{\rm ext}$
of the grains\footnote{Or rather of the quantity $Q_{\rm ext} / a$,
which is independent of the grain radius, $a$, in the small particle (Rayleigh) limit 
which is applicable in this context.} and the intrinsic density of the material. 
The parameters of the models discussed here can be found in Table 2.
Wind properties like the mass loss rate $\dot{M}$, the time-averaged
outflow velocity ${\langle u \rangle}$ and degree of condensation 
${\langle {f_{\rm c}} \rangle}$ are
direct results of the dynamical calculations.
All elemental abundances are assumed to be solar
except the one of carbon which is specified by an additional parameter, 
the carbon-to-oxygen ratio ${\varepsilon_{\rm C}/\varepsilon_{\rm O}}$.

\section{Results}

\begin{table}
\caption{Comparison of model results using different dust parameters.
         Model parameters: 
         luminosity $L_{\star}$ (in $L_{\sun}$), temperature $T_{\star}$ (in K),
         dust opacity data $\kappa_{\rm dust}$, intrinsic dust density 
          $\rho_{\rm dust}$ (in g/cm$^3$),
         mass $M_{\star} = 1.0 \,M_{\sun}$, carbon-to-oxygen ratio
         ${\varepsilon_{\rm C}/\varepsilon_{\rm O}} = 1.4$, pulsation period
         $P = 650$\,d, velocity of the inner boundary ${\Delta u_{\rm p}} = 4$\,km/s. 
         Results: mass loss rate $\dot{M}$ (in ${\rm {M_{\sun}} / yr}$),
         mean velocity at the outer boundary ${\langle u \rangle}$ (in km/s),
         mean degree of condensation at the outer boundary ${\langle f_{\rm c} \rangle}$.}

\begin{center}
 \begin{tabular}{|l|l|c|c||c|c|c|l|} \hline
$L_{\star}$ & $T_{\star}$ & $\kappa_{\rm dust}$& $\rho_{\rm dust}$ & $\dot{M}$ & ${\langle u \rangle}$ & ${\langle {f_{\rm c}} \rangle}$ & Comment \\
(L$_{\sun}$) & (K) & (cm$^{-1}$) & (g/cm$^3$) & (M$_{\sun}$) & (m/s) & & \\ \hline 
13000 & 2700 & J\"ager1000 & 1.99 &  $5.6 \cdot 10^{-6}$ & 15 & 0.05 & Model 1 \\
13000 & 2700 & J\"ager1000 & 2.25$^{*}$ & $7.3 \cdot 10^{-6}$ & 20 & 0.11 &  Model 2 \\
13000 & 2700 & Rouleau & 1.85 &  $4.3 \cdot 10^{-6}$ & 7 & 0.10 & Model 3 \\
13000 & 2700 & Rouleau & 2.25$^{*}$ &  $8.2 \cdot 10^{-6}$ & 18 & 0.31 & Model 4  \\
13000 & 2700 & J\"ager400 & 1.44 & - & - & - & Model 5 \\
13000 & 2700 & J\"ager400 & 2.25$^{*}$ & $2.1 \cdot 10^{-8}$ & 1 & 0.13 & Model 6 \\ \hline
10000 & 2600 & J\"ager1000 & 1.99 &  $7.0 \cdot 10^{-6}$ & 16 & 0.09 & Model 7  \\
10000 & 2600 & Rouleau & 1.85 & $2.3 \cdot 10^{-6}$ & 4 & 0.12 & Model 8 \\
10000 & 2600 & J\"ager400 & 1.44 &  - & - & - & Model 9 \\ \hline
\end{tabular} 
\end{center} 
{\it *} {\small Density of pure graphite.}
\end{table}
 
It is seen in Fig.\,1b that the new models coincide resonable well with observations of
 comparable stars. But at the same time it is clear from Table 2, that
the mean outflow velocity, ${\langle u \rangle}$, and the degree of condensation, 
${\langle {f_{\rm c}} \rangle}$,
change significantly with the dust data used. 


Comparing Model 1 and 3 the mean degree of condensation, 
${\langle {f_{\rm c}} \rangle}$, is much higher for the model using
the dust data with the lower opacity, but at the same time the mean
outflow velocity, ${\langle u \rangle}$, is higher for the model using
the dust data with the higher opacity. 


The degree of condensation also increases substantially if a higher intrinsic density for the 
material is assumed. 
In the models we have used both the true value of the material (Model 1,3,5,7,8,9) 
as it was determined in the laboratory as well as
the value of $\rho= 2.25$~g/cm$^3$, equivalent to the intrinsic
density of pure graphite (Model 2, 4, 6). The latter value has been used in many existing models 
(e.g.\ Fleischer et al. 1992; H\"ofner \& Dorfi 1997). 
The result of using the higher density of graphite instead of the right value, 
is that the models become much redder since more dust is formed. Even a small 
increase of about 10\% in the density of the dust material 
(as is the case from Model 1 to 2) 
results in a doubling of the degree of condensation and a substantial increase in 
the outflow velocity, ${\langle u \rangle}$. 
This stresses the importance of using the measured material value if possible,
since an other choice (even if it has been carefully considered) 
can create an artificial increase/decrease in the calculated mass loss of the models.

\acknowledgements

ACA greatfully acknowledges financial support from the Carlsberg Foundation. 
This work was supported by NorFA, the Royal Swedish Academy of Science and the Swedish 
Research Council.

\end{document}